\documentclass[lettersize,journal]{IEEEtran}
\usepackage{amsmath,amsfonts}
\usepackage{algorithmic}
\usepackage{algorithm}
\usepackage{array}
\usepackage[caption=false,font=normalsize,labelfont=sf,textfont=sf]{subfig}
\usepackage{textcomp}
\usepackage{stfloats}
\usepackage{url}
\usepackage{verbatim}
\usepackage{graphicx}
\usepackage{cite}
\usepackage[dvipsnames]{xcolor}
\usepackage{xspace}
\usepackage{paralist}
\usepackage{enumitem}
\usepackage{balance}
\usepackage{listings}
\usepackage{hyperref}
\newcommand{\eg}{\textit{e.g.}}
\newcommand{\ie}{\textit{i.e.}}
\newcommand{\etal}{\textit{et~al.}\xspace}

\newcommand{\tool}{Remex\xspace}

\newcommand{\re}[1]{\textcolor{black}{#1}}

\definecolor{codegreen}{rgb}{0,0.6,0}
\definecolor{codegray}{rgb}{0.5,0.5,0.5}
\definecolor{codepurple}{rgb}{0.58,0,0.82}
\definecolor{backcolour}{rgb}{0.95,0.95,0.92}

\lstdefinestyle{mystyle}{
    backgroundcolor=\color{backcolour},   
    commentstyle=\color{codegreen},
    keywordstyle=\color{black},
    numberstyle=\tiny\color{codegray},
    stringstyle=\color{codepurple},
    basicstyle=\ttfamily\tiny,
    columns=fullflexible,
    breakatwhitespace=false,         
    breaklines=true,
    breakindent=0pt,
    captionpos=b,                    
    keepspaces=true,                 
    numbers=left,                    
    numbersep=5pt,                  
    showspaces=false,                
    showstringspaces=false,
    showtabs=false,                  
    tabsize=2
}

\begin{document}

\title{Composing Data Stories with Meta Relations}

% \author{Anonymous Authors}

\author{
Haotian Li, Lu Ying, Leixian Shen, Yun Wang, 
Yingcai Wu,
and Huamin Qu
% % Leixian Shen~\orcidicon{0000-0003-1084-4912}, 
% % Haotian Li~\orcidicon{0000-0001-9547-3449}, 
% % Yun Wang~\orcidicon{0000-0003-0468-4043}, 
% % Tianqi Luo~\orcidicon{0000-0001-9547-3449}, 
% % Yuyu Luo~\orcidicon{0000-0001-9530-3327}, 
% % and Huamin Qu~\orcidicon{0000-0002-3344-9694}

\thanks{

% \item
H. Li, L. Shen, and H. Qu are with The Hong Kong University of Science and Technology.
E-mail: \{haotian.li, lshenaj\}@connect.ust.hk; huamin@cse.ust.hk
% \item

L. Ying and Y. Wu are with the State Key Lab of CAD\&CG, Zhejiang University, Hangzhou, China. 
E-mail: \{yingluu, ycwu\}@zju.edu.cn. 

Y. Wang is with Microsoft.
E-mail: wangyun@microsoft.com.

This paper is part of the thesis ``Bridging Data Analysis And Storytelling With Human-AI Collaborative Tools'' by the first author H. Li~\cite{li2024bridging}.
% \item
}}

\markboth{Journal of \LaTeX\ Class Files,~Vol.~14, No.~8, August~2021}%
{Shell \MakeLowercase{\textit{et al.}}: A Sample Article Using IEEEtran.cls for IEEE Journals}


\maketitle

\begin{abstract}
% \haotian{v3}
To facilitate the creation of compelling and engaging data stories,
AI-powered tools have been introduced to automate the three stages in the workflow: analyzing data, organizing findings, and creating visuals.
However, these tools rely on data-level information to derive inflexible relations between findings.
Therefore, they often create one-size-fits-all data stories.
Differently, our formative study reveals that humans heavily rely on meta relations between these findings from diverse domain knowledge and narrative intent, going beyond datasets, to compose their findings into stylized data stories.
% Differently, humans heavily rely on meta relations from diverse information behind datasets, 
% including domain knowledge and narrative intent, to compose these findings into stylized data stories, based on our formative study.
% Such a gap results in that AI-created data stories do not always reach humans' expectations.
Such a gap indicates the importance of introducing meta relations to elevate AI-created stories to a satisfactory level.
% Furthermore, from the formative study, we 
Though necessary, it is still unclear where and how AI should be involved in working with humans on meta relations.
% since they are often considered by human creators in an implicit and vague way.
% learned that meta relations are often considered by human creators but only in an implicit and vague way.
% Though , 
% To address the issue, we further investigates where and how AI should be involved to work with humans on meta relations.
% first investigated the commonly considered meta relations between data findings via a formative study identified two major sources of meta relations: domain knowledge and story's narrative intent. 
% In this paper, we explore how to augment AI-powered data storytelling with the consideration of meta relations.
% We first conducted a formative study to understand meta relations in human creators' practices and realized that meta relations are often considered but only in an implicit and vague way.
% The distillation and usage of meta relations highly depend on data workers' domain knowledge and their narrative intent for the data story.
% Due to the invisible and personalized characteristic, 
% % However, it is still 
% it is unclear where and how AI should be involved to work with humans on meta relations.
To answer the question, 
we conducted an exploratory user study with \tool, an AI-powered data storytelling tool that suggests meta relations in the analysis stage and applies meta relations for data story organization.
The user study reveals various findings about introducing AI for meta relations into the storytelling workflow, such as the benefit of considering meta relations and their diverse expected usage scenarios.
Finally, the paper concludes with lessons and suggestions about applying meta relations to compose data stories to hopefully inspire future research.

\end{abstract}


\maketitle

\section{Introduction}\label{sec:introduction}
\IEEEPARstart{D}{ata} stories appear everywhere from public communication~\cite{yang2023swaying}, business presentations~\cite{brehmer2021jam}, to personal reflection~\cite{kim2019dataselfie}.
They present multiple \textit{story pieces} supported by data with coherent \textit{relations} between them to convey messages between authors and creators~\cite{lee2015more, schroder2023telling}.
To facilitate data story creation, introducing artificial intelligence~(AI) assistance
has gained interest from both researchers and practitioners~\cite{li2023ai, chen2023does}.

When creating data stories, it is crucial to connect multiple data story pieces to convey a complete and clear message~\cite{li2023we}.
However, how to connect story pieces receives little attention in existing AI-powered data storytelling tools.
Following the seminal research in story piece organization~\cite{hullman2013deeper, kim2017graphscape}, most existing methods rely on data relations among story pieces to assist human creators.
For example, 
DataShot~\cite{wang2019datashot} groups data facts with similar data attributes or filters together and generates data posters accordingly.
Calliope~\cite{shi2020calliope} infers the relations, such as contrast and similarity, between story pieces based on their data attributes.
Then, it searches for the best sequence of these story pieces based on their relationship.
A more recent research, Socrates~\cite{wu2023socrates}  relies on the data relations to organize story pieces with consideration of users' feedback.
% \todo{However, data relations are not sufficient for AI to collaborate with humans effectively in organizing satisfactory data stories~\cite{li2023ai}.}
% due to the lack of alignment between AI and humans' consideration~\cite{li2023ai}.

To create coherent and clear data stories, humans often consider more factors beyond facts and relations from data, such as the social background, the goal of data stories, and also audiences' preference~\cite{burns2022invisible, li2023we, lin2023here}, which are overlooked by existing AI-powered tools.
As a result, the stories created by AI-powered tools lack the consideration of humans' preferences and expectations and are homogeneous when the data facts are similar.
Human creators still need to spend considerable effort in customizing them according to their requirements.
% Inspired by previous research~\cite{li2023we, burns2022invisible}, where the researchers identified that the meta information of data stories is crucial for authoring and understanding them, 
To fill the gap, we introduce \textit{meta relations} to reflect the connection between story pieces more comprehensively.
Different from data relations that can be inferred from story pieces solely, meta relations describe the relations that are from the meta information of data stories, such as background information of these stories and analytical questions that the data story aims to answer.
Incorporating such relations can enable AI to understand human creators' mindsets and, therefore, organize data stories closer to their expectations.
% and boost the collaboration between human creators and AI.
% As a result, they have the potential to better 
% Relations from meta infromation
% Considering such relations can enable

An example of considering meta relations in data storytelling is from an award-winning data article ``Scientific Proof that Americans are Completely Addicted to Trucks'' from Bloomberg, where the auto sales in 2014 are introduced~\cite{pearce2015car}.
% \footnote{The data article receives \href{https://www.informationisbeautifulawards.com/showcase/801-scientific-proof-that-americans-are-completely-addicted-to-trucks}{Kantar Information is Beautiful Awards 2015}.} 
The article introduces the fact of the decreasing sales of the hybrid electric car model, Toyota Prius, and the facts of the increasing sales of other plug-in and pure electric car models, including Nissan Leaf, Tesla S, etc., consecutively.
Since meta relations about these car models, \ie, \textit{they belong to different electric car categories with competitive natures}, are considered, the article compares these car models to convey the insightful message of a growing interest in plug-in and pure electric cars of the market.
However, if considering the facts solely or referring to the dataset of the article shown at the bottom, we notice that these cars do not share any commonalities in categorical features, indicating a weak data relation among them.
When only such relations are considered, these facts are unlikely to be introduced together.
The example clearly demonstrates that meta relations between data facts are indispensable to a well-organized data story.
Such findings motivate us to explore how to introduce meta relations to AI-powered data storytelling tools to compose stories that align with humans' knowledge and preferences.

% \begin{figure}
%     \centering
%     \includegraphics[width=\linewidth]{figures/motivating_example.pdf}
%     \caption{A motivating example to showcase the necessity of meta relations in data storytelling from Bloomberg.}
%     \label{fig:motivating_example}
% \end{figure}

\begin{figure}
    \centering
    \includegraphics[width=\linewidth]{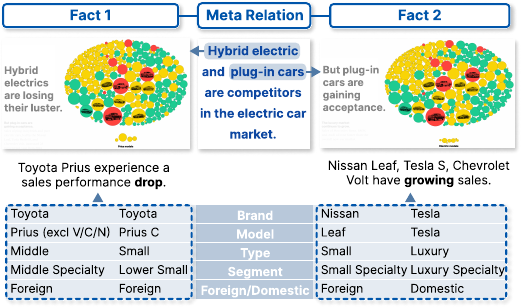}
    \caption{A motivating example to showcase the necessity of meta relations in data storytelling from Bloomberg~\cite{pearce2015car}.}
    \label{fig:motivating_example}
\end{figure}

In our research, we first conducted a formative study to enhance our understanding of meta relations in practice.
From these responses, we summarized findings about meta relations, including the categorization of diverse meta relations based on the source types (\ie, domain knowledge and narrative intent) and their challenges with meta relations. 
% and the efforts in distilling them from domain knowledge.
These lessons further informed us of the considerations in developing an AI-powered tool for composing data stories with meta relations, such as allowing free-form meta relation definition and assisting users in identifying meta relations.
% Following the consideration, we designed a large language model (LLM)-driven approach to automatically identify meta relations between data facts and compose data stories for 
% Based on the design considerations
However, it is still unclear where and how humans would like to receive assistance from AI on meta relations in data story creation workflow.
\re{To figure out the issue, we developed a prototype tool, \tool, and conducted an exploratory user study with it.}
\tool enables inferring meta relations between story pieces automatically for users' selection and then organizing story pieces into data stories with joint consideration of meta and data relations automatically.
The participants in the user study were instructed to create data stories with both \tool and Notable~\cite{li2023notable}, another notebook tool that supports data storytelling based on data relations alone, without considering meta relations.
% Their feedback verified the usefulness of considering meta relations explicitly in \tool.
According to participants' feedback, we summarized a series of findings about applying meta relations in data storytelling.
First, they confirmed the usefulness of introducing meta relations into the storytelling workflow explicitly, such as providing hints about the direction of data stories and connecting story pieces more tightly.
Second, we summarized various desired usage of meta relations, like showing underlying relations between two selected facts and providing suggestions for next story pieces.
Furthermore, we also noticed challenges in delivering meta relations, such as displaying meta relations with minimized mental load.
% Their feedback not only verified the usefulness of considering meta relations explicitly in \tool but also revealed a series of findings about applying meta relations in data storytelling, such as the diverse usage scenario of meta relations in the data storytelling lifecycle.
% Then they were interviewed about experiences with two tools.
% From their feedback, we summarize a series of findings about applying meta relations in data storytelling, such as the diverse usage scenario of meta relations in data storytelling lifecycle.
We also provide lessons learned in the research and suggestions for future AI-powered tools that aim to incorporate meta relations.

To conclude, our contributions include:
\begin{compactitem}
    % \item \textbf{proposing meta relations} as a new perspective to reflect the connections between story pieces.
    \item \textbf{introducing meta relations} to enhance the connections between story pieces and facilitate automatic data storytelling.
    \item \textbf{unveiling meta relations and pain points associated with them} through a formative study with data workers.
    \item \textbf{summarizing design considerations} for tools that integrate meta relations into the storytelling workflow and exemplify the considerations with a proof-of-concept tool, \tool.
    \item \textbf{outlining the benefits and usage scenarios of meta relations} based on a user study where participants have hands-on experiences to compare \tool with a baseline without meta relations.
    \item \textbf{summarizing the lessons} for researchers to enhance the consideration of meta relations in future studies.
\end{compactitem}

\section{Related Work}
We review the prior literature from three perspectives: AI-powered data storytelling tools (Sec.~\ref{sec:ai-storytelling}), automatic data story piece organization (Sec.~\ref{sec:auto-organization}), and meta information in data storytelling (Sec.~\ref{sec:meta-information}).
% and data storytelling in computational notebooks (Sec.~\ref{sec:storytelling-notebook}).

\subsection{AI-powered Data Storytelling}\label{sec:ai-storytelling}
Telling coherent and compelling data stories often requires diverse skills, such as visualizing data and drafting scripts, and considerable human efforts due to its long pipeline~\cite{lee2015more, chevalier2018analysis}.
To democratize data storytelling, researchers and practitioners have explored how to leverage the power of AI techniques in the whole workflow~\cite{chen2023does, li2023we}.

Recent research has proposed to characterize existing AI-powered data storytelling tools according to the stages where the tools facilitate, including analysis, planning, implementation, and communication~\cite{li2023we}.
The tools enhancing the \textit{analysis} stage mainly mine data facts from datasets to assist human creators or provide materials for story generation.
For example, Shi~\etal~\cite{shi2020calliope} and Lu~\etal~\cite{lu2021automatic} generate data stories in the scrollytelling form based on mined data facts in the analysis stage.
Erato~\cite{sun2022erato} and Notable~\cite{li2023notable} suggest potential data facts for humans' selection.
In the \textit{planning} stage, the data story creators need to consider how to arrange the story pieces about data findings into a coherent story.
Representative tools that cover this stage include DataShot~\cite{wang2019datashot} and Calliope~\cite{shi2020calliope}.
Sec.~\ref{sec:auto-organization} provides a more comprehensive introduction to how these tools organize data story pieces.
The \textit{implementation} stage has received the most attention from researchers.
The tools often assisted in creating static or dynamic visualizations.
For example, 
DataQuilt~\cite{zhang2020dataquilt} extracts reusable visual elements with AI techniques to facilitate new infographic creation.
Text-to-Viz~\cite{cui2019text} generates infographics according to users' natural language descriptions.
More recently, Ying~\etal~\cite{ying2023reviving} and Shen~\etal~\cite{shen2023data} proposed approaches to generate data videos automatically from static charts with assistance from LLMs.
Only a few tools are proposed to enhance the \textit{communication} phase where data stories are directly communicated with the audiences.
Lee~\etal~\cite{lee2013sketchstory} and Hall~\etal~\cite{hall2022augmented} facilitate more vivid communication of data stories through recognizing storytellers' gestures and presenting corresponding story pieces automatically.

% The tools to facilitate the planning stages are the most closely related to our 

Following the AI-powered data storytelling trend, to enable AI to better understand the intentions of human creators, this paper goes beyond data-level information and explores the integration of meta relations (e.g., domain knowledge and story's narrative intents) beyond datasets.
As part of our research, we developed a proof-of-concept tool called \tool, which incorporates meta relations into the workflow of communicating data findings in notebooks with slides, covering analysis, planning, and implementation stages and emphasizes how to organize data stories with meta relations in the planning stage more conveniently and effectively. In the next section, we will introduce how previous tools achieve automatic data story organization and the aspect that meta relations would like to improve.

% \haotian{to be updated} In our paper, we propose \tool, an AI-powered data storytelling tool, to facilitate communicating data findings in notebooks with slides. Specifically, \tool covers analysis, planning, and implementation stages and emphasizes how to organize data stories with meta relations in the planning stage more conveniently and effectively. In the next section, we will introduce how previous tools achieve automatic data story organization and the aspect that \tool would like to improve.

% assists users in identifying and documenting data findings in the analysis stage, achieves an automatic data story organization with considering the meta relations and data relation

\subsection{Automatic Data Story Piece Organization}\label{sec:auto-organization}
% Organizing data story pieces is crucial for conveying a clear message with a coherent data story.
When authoring data stories with multiple data facts, it is necessary to organize these data facts coherently~\cite{li2023we}.
Previous research has identified various perspectives to be considered when organizing data stories, such as narrative strategies~\cite{yang2021design}, audiences' perceived mental load~\cite{hullman2013deeper, kim2017graphscape}, and the meta information of the story~\cite{li2023ai, burns2022invisible}.
It is demanding to consider these factors comprehensively.
To address the challenge, automatic approaches have been proposed.

As pioneer research in automating data story organization, Hullman~\etal~\cite{hullman2013deeper} identified a set of common connection patterns between adjacent charts in data stories, mostly from the data perspective, such as dimension and measure changes.
Then they proposed a graph-based approach to arrange charts based on the relations between their data attributes to minimize audiences' transition costs.
Kim~\etal~\cite{kim2017graphscape} proposed GraphScape to enhance Hullman~\etal by enriching the data relations and developing a workable model based on Vega-Lite~\cite{satyanarayan2016vega}.
Following their approaches, later research still relies on data relations to organize story pieces automatically.
DataShot~\cite{wang2019datashot} organizes data facts into groups based on their shared topics, such as common filters.
Calliope~\cite{shi2020calliope} proposes a group of data relations between data facts, including contrast and similarity, and organizes data stories based on such relations with a Monte Carlo Tree Search.
A similar approach is adopted by Socrates~\cite{wu2023socrates} with an additional consideration of humans' feedback. 
Erato~\cite{sun2022erato} augments Calliope by allowing human creators to fix an important fact sequence first and then fill in additional facts to enhance humans' agency in automatic data story organization.

Comparing the factors considered by existing approaches and humans' practices, we conclude that the previously proposed approaches have a common drawback where they consider merely objective and inflexible data relations.
They can result in ``one-size-fits-all'' data stories and fail to cater to users' specific needs according to their backgrounds or preferences, as indicated in previous research~\cite{li2023notable, li2023we}.
% Such an issue can result in unsatisfactory data stories generated by these tools, as indicated in previous research~\cite{li2023notable, li2023we}.
% \haotian{one size fits all}
% \haotian{why unsatisfactory? what are the benefits by meta relations}
To take one step further, we explore how to incorporate more flexible and customized meta relations from metadata of datasets to provide effective assistance to story piece organization automatically.

\subsection{Meta Information in Data Storytelling}\label{sec:meta-information}
\re{Our paper introduces meta relations as a new perspective to delineate the connection between data facts.
They are extracted from meta information of data stories, \ie, domain knowledge and narrative intent.
The introduction of meta relations is motivated by previous research about meta information in data storytelling.}

\re{According to Chevalier~\etal~\cite{chevalier2018analysis}, metadata was necessary for data journalists to understand the data before making stories.
Hullman and Diakopoulos~\cite{hullman2011visualization} argued that the narrative intent of visualization creators affected the final data story.
Li~\etal~\cite{li2023ai} indicated that the meta information of datasets, such as social context and project background, was often considered when making data stories.
Furthermore, AI assistance for meta information collection in data storytelling is desired.
Besides the importance when making stories, meta information can also affect data story comprehension.
Burns~\etal~\cite{burns2022invisible} found that the metadata of communicative visualizations (\eg, dataset descriptions and the goal of visualizations) could enhance the perceived thoroughness of readers and help them assess whether the visualization is trustworthy.
Lin~\etal~\cite{lin2023here} mentioned that domain knowledge of datasets could affect the understanding of data insights.}

\re{These studies inform that meta information is crucial for both creating and understanding data stories, though it was seldom considered in previous research on AI-powered data storytelling.
In our paper, we propose meta relations as a potential way to enhance the consideration of meta information in data story creation.
Our study aims to understand the practices and expected AI assistance for meta relations.
}

\section{Formative Study} \label{sec:formative}
To learn about the practices, pain points, and expectations of meta relations, we conducted a formative study with six data workers.
\re{In our paper, data workers are those who conduct data science work in daily jobs regardless of whether they are professionals or amateurs~\cite{zhang2020data, muller2019data}.}

% Age = 27, 26, 26, 25, 27, 27
% Experience = 7.5, 5, 8, 1, 4.5, 5

\subsection{Participants}
\re{We invited six participants (4 males, 2 females, $Age_{mean}=26.33$, $Age_{std}=0.75$, denoted as P1-P6) from various disciplines, including computer science (P1, P2, P5), information design (P3), economics (P4), and chemistry (P6) through our professional network.
We recruited participants with diverse backgrounds to enhance our understanding of meta relations with multiple viewpoints from various data workers.}
They have rich experience in analyzing and communicating data ($Experience_{mean}=5.17~years$, $Experience_{std}=2.29~years$).
% , with convenient sampling.
% They all frequently analyze and communicate data (\ie, ranging from every day to 1-3 times per month) with at least one year of experience.

\subsection{Procedure}
% All studies were conducted via one-on-one online or offline meetings.
Before the study, we first briefly introduced the purpose of the research.
After collecting the consent of recording and using participants' demographic information and opinions expressed in the study, we first enquired about the participants' background information, such as their ages, genders, and work nature.
Then we started our study with a brief introduction to data storytelling, where we explained the terms such as story pieces and meta relations.
Next, the participants were asked to share their practices in identifying and applying meta relations when communicating data findings.
% Specifically, we ask the
To help them recall the experiences, we also designed a task where the participants were instructed to compile a data story to introduce SUV car sales.
Specifically, we provided twelve data facts about SUV cars from the widely applied car sales dataset in previous data storytelling research~\cite{wang2019datashot, li2023notable}.
The dataset includes five attributes: car models, brands, categories, sales, and years.
We also provided background information about the SUV models from Wikipedia.
They were instructed to identify meta relations between 12 data facts and organize them into a satisfactory data story using around 15-20 minutes.
During the study, they were encouraged to express how they decided on the story organization.
After finishing the task, we asked them to introduce the story and meta relations inside.
Finally, we asked the participants about their challenges with meta relations and what assistance they expect to enhance telling data stories with meta relations.
\re{The study with one participant lasted for around 40 minutes.
The whole study was recorded and transcribed.
Following the analysis approach in previous research~\cite{li2023notable}, after interviews, the first author organized the participants' opinions based on the notes and recordings.
Then, the results were discussed with co-authors to reach a consensus.
Finally, we identified two key findings from the study as below.}

\subsection{Results}
% Based on participants' opinions and our observations on users' actions in the study
\re{Based on participants' opinions}, we summarized two key results about the practices regarding meta relations.
% in practices.

\textbf{The types of meta relations are diverse and are mostly identified based on their sources.}
Overall, all participants confirmed that they often took meta relations into consideration when performing analysis and follow-up communication.
When introducing the meta relations they often consider, the participants often categorize them based on the sources of meta relations.
% According to their descriptions, we realized that most
% P1 and P4 mentioned that they would consider two types of meta relations based on analytical knowledge and domain knowledge, respectively.
We notice that there are two types of sources from which meta relations come, including \textit{domain knowledge} (P1, P2, P3, P4, P5, P6)  and \textit{narrative intent} (P3, P5).
% The last type is from the visual aspect, where the \textit{visual presentation} of story pieces (P5) is considered.

\textit{Domain knowledge} refers to the expertise related to a specific domain that is not included in the dataset, \eg, the origins of cars in the car sales dataset.
Domain knowledge is the most commonly mentioned type of meta relation source.
For example, P6 organized the facts in our task using the size of SUV cars (\eg, mid-size and full-size) based on his knowledge.
Similarly, P5 considered BMW models to be representative of luxury models and compared them with others.
These cases are mainly domain-related factual knowledge following the categorization of knowledge in Bloom's taxonomy~\cite{forehand2005bloom}, where knowledge is categorized into four types: factual, conceptual, procedural, and meta-cognitive, following the order of difficulties in mastering.
We also noticed the existence of conceptual domain knowledge, which indicates how domain principles are applied to connect facts from data.
P4, as an expert in economics, pointed out that conceptual domain knowledge plays an important role in organizing their findings.
P4 mentioned that they often need to consider meta relations based on their models and theories.
He often picks important data attributes to analyze and present based on the factors considered in previous theoretical models.
The opinion is echoed by P6.
Beyond conceptual knowledge, our participants also leverage procedural knowledge about the domain to build the meta relations following their analytical pipelines.
For example, P1 mentioned that he arranged the data facts based on his commonly used strategy, where the story pieces are introduced according to the car models first and then the temporal analysis.
P4 also introduced an example where they leveraged their expertise to link the facts of the fast-increasing consumption of electricity, the relevantly slowly increasing carbon dioxide density in the atmosphere, and the increasing investment in green power to showcase the impact of such investment.
We did not notice any meta-cognitive knowledge since it is about how other knowledge is learned and might be beyond our scope.

\textit{The narrative intent} mainly includes the question that the story aims to answer and also the goals of the story.
P3, as a designer, highlighted the impact of data story goals.
She mentioned that the take-home message would affect how she organized story pieces.
In our task, she noticed that two SUV models, BMW X5 and X6 had increasing sales between 2007 and 2011 while the other car models experienced a sales drop.
She said that she would introduce BMW models first if the key message is that the overall SUV market is depressed.
On the contrary, if the story's goal is to highlight the business success of BMW models, it is better to introduce the unsatisfactory performance of other models first to highlight BMW's unusually increasing sales.

% The \textit{visual presentation} of story pieces highlights a unique perspective of considering meta relations in data storytelling.
% P5 pointed out that visual representations may also affect how he would organize the story pieces.
% For example, if two facts share the same number, \ie, a car model's sales over the year are 18,000 while another model's minimum sales over the year are also 18,000.
% Though they may not have strong data relations, they are likely to be put together to create a consistent visual experience and emphasize that the second car model has much better sales.
% \haotian{@Lu, please check and add more details}

To summarize, we have two observations about meta relations. 
First, unlike data relations, \textit{meta relations have a more flexible and ambiguous nature based on users' intents and preferences}.
As a result, meta relations may not have a concrete and complete categorization of their detailed meaning, such as contrast or similarity in data relations~\cite{shi2020calliope}. 
Second, though a detailed categorization of meta relations is not feasible, \textit{our participants unanimously described the classification of meta relations using the sources where meta relations are identified}.
% using their sources}.
Specifically, we identify two common sources of meta relations in data stories, \ie, domain knowledge and narrative intent.

\textbf{Applying meta relations is non-trivial and therefore AI assistance is expected.
}
Though meta relations are widely applied, it is non-trivial for our participants to apply them in data storytelling.
We noticed that most of the challenges are related to domain knowledge.

Several participants mentioned that their challenges are related to 
% domain knowledge (P1, P2, P4).
% P2 mentioned his challenges in 
acquiring and digesting domain knowledge to distill meta relations (P1, P2, P3, P5, P6).
When P2 conducted the tasks in our formative study, to help himself understand the content about SUV background, he fed all materials to ChatGPT, a large language model (LLM)-driven chatbot, and asked it to summarize them into bullet points.
In this way, he could easily understand the differences and commonalities between SUV models.
P1, P3, and P5 also found domain knowledge is often overwhelming and could be hard to digest without any assistance, which is described as \textit{``explicit hint''} on potential meta relations by P1.
This finding aligns with the common practices of data workers where they often work with diverse domain experts in different projects~\cite{li2023ai, piorkowski2021ai, lin2023here}.
They rely on domain knowledge provided by these experts to make sense of data and interpret the findings from datasets. 
P2 also mentioned the difficulty of collecting domain knowledge by talking about his experience with a mobile game-related analysis project.
He said \textit{``I spent five months playing the game!''} to collect the domain knowledge about the game mechanism for the analysis task.
P4 also indicated his challenges with conceptual knowledge in the domain.
He often needed to search for literature to support and verify the linkage between his results and models, where assistance was also expected.
Besides domain knowledge, P3 mentioned that she might be unsure how to better compile a data story following her narrative intent.
As a result, she would like to ask AI for assistance in brainstorming potential ways to tell stories or assessing the pros and cons of different approaches.
% \haotian{@Lu, please add more details}

From the participants' opinions, we realized that \textit{the main challenge lies in applying meta relations from domain knowledge} due to the unfamiliarity of a specific domain.
% We consider that knowledge might be more familiar to data workers considering their background while they often know less about domains.
They were keen on having approaches to help them digest domain knowledge and provide hints in potential meta relations.
Furthermore, \textit{the application of AI in addressing their challenges is observed}.
P2 directly applied ChatGPT to our task, while P3 also mentioned the possibility of introducing AI to her workflow.
Such observation implies the potential of introducing powerful AI models to collaborate with humans for the aforementioned issues.

\section{Design Considerations}
According to the literature and our formative study, meta relations are important in organizing story pieces into a personalized and vivid data story.
\re{This section summarizes considerations for incorporating meta relations in AI-powered data storytelling tools based on their definition.}

We first formalize the definition of meta relations.
In previous research~(\eg, \cite{hullman2013deeper, kim2017graphscape, li2023notable}), a data relation often includes four components: two story pieces connected by the relation, the type of the relation, and a score to delineate the importance of the relation.
To keep consistent with data relations, we define a meta relation $relation_{AB}$ from $fact_A$ to $fact_B$, with a relation type and an importance score of the relation, as a quadruple, $relation_{AB}:= (fact_A, fact_B, type, score)$.
\re{In the definition, $fact_A$ and $fact_B$ are two data facts (\eg, ``Toyota Prius experience a sales performance drop'' and ``Nissan Leaf, Tesla S, Chevrolet
Volt have growing sales'' in Fig.~\ref{fig:motivating_example}).
$type$ is a concise description of a meta relation between two facts using the meta information of the dataset (\eg, ``Hybrid electric and plug-in cars are competitors in the electric car market'' in Fig.~\ref{fig:motivating_example}).
$score$ quantifies the importance of the meta relation with a score between 0 and 1, where a larger value indicates a higher importance.}
Notably, there can be multiple relations between $fact_A$ and $fact_B$ with different types and corresponding scores.

The considerations include four aspects: 
how tools handle the definition of meta relations, what functions should be included, and what techniques could be applied.
The first two considerations \textbf{C1}-\textbf{C2} elicit \textit{the requirements on meta relation definition}.

\textbf{C1.} \textbf{Defining meta relation types flexibly with natural language.}
The study results reveal an important characteristic about meta relations: \textit{they depend on the detailed meta information and thus are flexible}.
Based on the finding, it is almost impossible to categorize meta relation types into a fixed number of classes, which is common in identifying data relations with AI~\cite{shi2020calliope}.
% Therefore, we did not follow the approach of data relations, where a set of relation types is well-defined.
As a result, \textit{the types of meta relations should be described with free-form natural language}, inspired by a recent research~\cite{yan2023xcreation}.
In this way, it is possible to fully unlock the potential of meta relations as a medium to convey human creators' intended relations between story pieces to AI modules in these tools.

\textbf{C2.} \textbf{Estimating the importance of meta relations in a universal way from multiple angles.}
When applying meta relations in composing data stories, the importance scores can be essential references for ranking multiple relations between two facts.
% Such ranking can help users to decide whether a meta relation should be considered.
Previously, the importance scores of data relations are often decided with specific criteria according to the relation type, such as the overlapping between attributes in adjacent facts~\cite{li2023notable}.
However, it is non-trivial to use a similar approach to compute scores for meta relations due to their flexibility, as indicated in C1.
% What makes the task challenging is the diverse sources of meta relations, including domain knowledge and narrative intent.
% The importance of a meta relation can be affected by both domain knowledge and narrative intent.
Therefore, it is essential to design a universal approach to quantify the importance of meta relations with considering the sources of meta relations concurrently.

% \vspace{1em}
% \textbf{C1}-\textbf{C2} discuss how the flexible characteristics of meta relations may affect handling the definitions of meta relations in data storytelling tools.
Next, \textbf{C3}-\textbf{C5} present \textit{considerations about detailed functions that tools should include to address users' pain points}.
% \vspace{1em}

\textbf{C3.} \textbf{Suggesting potential meta relation from users' provided domain knowledge automatically.}
% Previous empirical research points out that data workers may not have sufficient domain knowledge so they often need to refer to domain experts or other meta information sources to support their data analysis and storytelling~\cite{lin2023here, li2023ai}.
Based on our study results and prior literature~\cite{lin2023here, li2023ai},
data workers may need to spend considerable time and effort to provide meta relations manually.
Furthermore, the context switch between data analysis and information seeking is often undesired~\cite{li2023notable}.
We envision that it is essential to provide an efficient way to assist data workers in distilling the meta relations between data facts from the documents provided by data workers. 
Inferring relations from provided documents could limit the scope of meta relation inference and ensure that the inferred ones match expectations.

\textbf{C4.} \textbf{Following users' diverse narrative intent in the workflow.}
As revealed in our formative study, diverse narrative intent often guides organizing story pieces into a coherent story, such as the analytical question and the intended take-home message.
Therefore, these tools should allow users to express their narrative intent freely and consider such intent
% After obtaining the narrative intent, these tools should take narrative intent into consideration 
when suggesting meta relations and organizing story pieces.

\textbf{C5.} \textbf{Providing users with sufficient flexibility to select, modify, and delete automatically generated meta relations and provide customized meta relations. }
Compared to data relations, meta relations can be more diverse and dependent on the provided documents (\textbf{C1}).
Therefore, the inferred meta relations may not be fully aligned with data workers' knowledge and expectations.
It is crucial to provide convenient interfaces and interactions for data workers to select useful meta relations, modify their details, or reject those incorrect and redundant ones.
Also, extra meta relations should be easily added.

\textbf{C6.} \textbf{Supporting meta relations in the entire lifecycle of data storytelling.}
Considering the close connection between all stages when creating data stories~\cite{lee2015more}, the support of meta relations should also be reflected in multiple stages.
% such as assisting  the identification of meta relations in the analysis stage and the organization of story pieces based on meta relations.

To achieve \textbf{C1}-\textbf{C5}, there are several technical barriers.
First, considering the free-form expression of meta relations (\textbf{C1-C2}), \textit{we can hardly design a fixed set of algorithms to decide the relation type and compute the relation score} when recommendation (\textbf{C3}).
Second, when organizing data facts based on meta relations, \textit{the AI should be able to interpret the semantic meaning of both these free-form meta relations and the narrative intent} (\textbf{C4}-\textbf{C5}).
Third, \textbf{C6} requires the AI system should have the ability to care for different tasks in the whole workflow.
Lastly, since there is no pre-defined set of meta relation types, \textit{it is hard to collect data and train an AI model to fulfill the needs}.

% \vspace{1em}

\textbf{C7.} \textbf{Leveraging LLMs to suggest and interpret meta relations carefully.}
The emerging large language models (LLMs) shed light on addressing the challenges due to their advanced performance in understanding and generating natural language and the zero-shot learning ability through prompt engineering~\cite{li2023we}.
Such features allow LLMs to distill meta relations from domain knowledge and organize data facts without additional training.
% Therefore, LLMs can be leveraged in meta relation-related algorithms.
On the other hand, their application should be careful due to their drawbacks, \eg, delay and unreliable results.
% mitigated during the design of LLM-based systems.

% \vspace{1em}

% \haotian{to be added}
% Considering the popularity of computational notebooks as the data analysis platform of data workers~\cite{rule2018exploration}, we decide to design \tool as a computational notebook extension to facilitate a seamless experience from analyzing to communicating data.
% Users can finish the analysis, planning, and implementation stages in the data storytelling workflow within their familiar notebook interface.
% According to previous research~\cite{li2023notable}, notebook users expect to finish the whole lifecycle from data analysis, finding organization, and slide creation in notebook interfaces seamlessly.
% Following their observation of the pain points in data storytelling, \tool provides assistance in three tasks to reduce users' workload: documenting data findings, organizing them into a coherent sequence, and creating a preliminary version of slides for storytelling.
\section{\tool: Exemplifying the Design Considerations}\label{sec:computational}
% The formative study informed a set of design considerations for considering meta relations in AI-powered storytelling tools.
% However, we realized that there is still a critical decision: 
% Besides the considerations above, we conducted an exploratory user study to figure out where and how AI should intervene when telling stories with meta relations.
To explore where and how AI should intervene when telling stories with meta relations, we designed a tool, \tool (\textbf{RE}lax the constraints of composing data stories with \textbf{ME}ta relations), that follows the design considerations.
\tool integrates AI-powered computational modules to support the lifecycle of storytelling in the computational notebook, a common platform to analyze and communicate data~\cite{rule2018exploration} (\textbf{C6}).

\subsection{Overview}
\begin{figure}
    \centering
    \includegraphics[width=\linewidth]{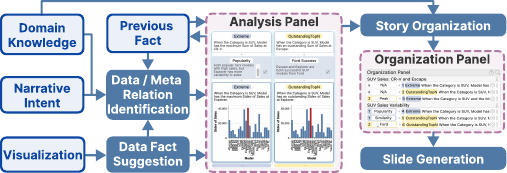}
    \caption{This figure illustrates the structure of \tool. More details about the analysis and the organization panels are available in Fig.~\ref{fig:interface}.}
    \label{fig:workflow}
\end{figure}

% \tool is a data storytelling tool that emphasizes distilling and applying meta relations for data workers to 
% organize their findings into a coherent data story to fulfill their diverse needs.
As Fig.~\ref{fig:workflow} shows, \tool is composed of five computational modules powered by AI techniques working at the backend and two interactive modules that allow humans to view and improve the output of computational modules (\textbf{C5}).
Users can provide the story-related domain knowledge and their narrative intent in the text format when starting the data analysis session in computational notebooks.
Then, users can add visualization specifications in Vega-Lite~\cite{satyanarayan2016vega} in each cell for \tool to illustrate potentially interesting data facts using the \textit{data fact suggestion module} and suggest meta relations with previously user-interested facts with the \textit{meta relation identification module}.
Besides, \textit{data relation identification module} complements the meta relation identification module to infer data relations for the follow-up organization.
The facts and meta relations are then presented in the \textit{analysis panel} for users' selection and modification.
When users feel satisfied with a fact with or without meta relations, they can add them to their slides for communication purposes.
The fact will be inserted into a draft slide deck by the \textit{story organization module}.
The slides are shown in the \textit{organization panel} for users' adjustment.
When the analysis ends, the slide deck can be exported for follow-up improvements or direct presentation with the \textit{slide generation module}.
% Considering the purpose of \tool is to realize our goal of introducing meta relations into storytelling, we focus on the design of meta relation-associated modules in Secs.~\ref{sec:meta_identification}-~\ref{sec:story_organization} and interactive modules in Sec.~\ref{sec:interactive_modules}.
% The details about other modules can be found in our supplemental materials.

\subsection{Data Fact Suggestion}
\re{\tool leverages a data fact suggestion module to lower users' workload in documenting data findings when they realize any notable findings.
% They can directly add the facts to their data story with a simple selection of data facts instead of manually typing the findings' descriptions. 
The module applies the well-established fact mining algorithm~\cite{ding2019quickinsights,wang2019datashot,li2023notable} to infer data facts from user-specified visualizations with seven dimensions, including a set of filters to describe how the data subset is derived (\textit{subspace}), the independent and dependent variables in a chart (\textit{dimension} and \textit{measures}), and four attributes to describe the data fact (\ie, the \textit{type} of the data fact, \textit{parameters} to indicate the fact details, the \textit{focus} data point of the fact, and \textit{scores} to delineate a fact's importance and alignment with users' interest).}
% subspace to record the filters that are applied to gain the
% visualized data subset; measures to indicate the dependent variables
% in the chart; dimension to represent the independent variable in
% a chart; type of data facts (e.g., trend and outlier); parameters to
% describe the details of the fact, such as the direction of trends; focus
% to document the data point that is emphasized in the fact; and score
% to measure both how the fact matched users’ intent of exploration
% and how important the fact is.
\re{The output of this module is data facts with text descriptions and visual illustrations.}

\subsection{Data Relation Identification}
% Once the data facts inside a newly specified visualization are recommended by the data fact suggestion module, 
% After obtaining suggested data facts inside a newly specified visualization, the data and meta relation identification modules compute their relations with previously selected facts by users.
\re{The data relation identification module computes two types of data relations between data facts with the algorithm in a previous research~\cite{li2023notable}.}

\re{Following the principles of sequencing story pieces~\cite{hullman2013deeper}, the first group of relations delineates the similarity between data facts through the overlap between data facts.
% Such relations help arrange facts with a higher similarity to close positions in the story piece sequence to minimize transition costs.
The detailed relation types include the overlapping between subspaces, measures, dimensions, focuses, and fact types.
Furthermore, the scores in such overlapping relations reflect how large the overlapping between two facts' certain attributes is.
We compute it as the Intersection-Over-Union (IoU) score~\cite{rezatofighi2019generalized} between two facts' specific attributes, such as subspaces or measures.}
% For example, suppose one fact $C$ has a subspace defined by \textit{``Brand=BMW''} and \textit{``Category=SUV''} and another fact $D$ is with \textit{``Brand=Honda''} and \textit{``Category=SUV''}, the relation about their overlapping of subspace is $relation_{CD} = (C, D, overlapping_{subspace}, 0.3333)$.

% Besides, another group of data relations facilitates the decision of the detailed sequence of similar facts.
\re{The second group of relations delineates the temporal relationships between two facts and the relationships between two facts' importance scores.
The temporal relationships include the temporal order of two facts' subspaces and focus. 
The rationale behind considering them is to introduce two facts in chronological order.
The introduction of the relationships between the two facts' strengths aims to arrange the important facts before the less important ones.
For these relations, we consider the relation strength score as a binary value, \ie, 0 or 1.
When the two facts in the relation follow the chronological order or the more important fact is before the less important face, the relation score is 1.
Otherwise, the score is 0, or, in other words, the relation does not exist.}
% For example, suppose one fact $E$ has a focus \textit{``Year=2000''} and the focus of another fact $F$ is  \textit{``Year=2010''}, the relation about the temporal order of focus is $relation_{EF} = (E, F, temporalOrder_{focus}, 1)$ while $relation_{FE}$ is invalid.

\subsection{Meta Relation Identification}\label{sec:meta_identification}
\re{According to our formative study, AI assistance is desired by some data story creators to distill meta relations between data facts from domain knowledge (\textbf{C3}).
At the same time, the AI should consider how the meta relation may fulfill the user's narrative intent (\textbf{C4}).
We design a meta relation identification engine to help users discover meta relations.}

\begin{figure}
    \centering
    \includegraphics{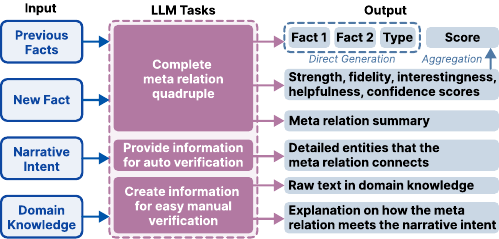}
    \caption{The figure explains the meta relation identification module.}
    \label{fig:meta_llm}
\end{figure}

Following \textbf{C7}, we leverage LLMs to identify flexible meta relations and infer relation importance scores.
Informed by their zero-shot learning ability~\cite{wei2021finetuned}, we prompt LLMs to finish three tasks, including completing the quadruples of meta relations and supporting automatic and manual verification, as shown in Fig.~\ref{fig:meta_llm}.
We carefully design the two verification-related tasks to 
minimize unreliable results, such as hallucinated response~\cite{lee2023benefits}, in the completion task.
We feed both previously selected and new data facts, narrative intent, and domain knowledge into LLMs to conduct these three tasks.

The first task asks LLMs to \textit{complete the quadruples of meta relations}, \ie, two facts, their relation type, and the relation score.
The model directly identifies the two facts and their meta relation type as a natural language description from domain knowledge  (\textbf{C1}), as well as a summary for human creators to quickly understand the meta relation in one glance.
Furthermore, to compute the relation score as comprehensively as possible (\textbf{C2}), we ask LLMs to self-rate the meta relation from five perspectives: \textit{strength}, \textit{fidelity}, \textit{helpfulness}, \textit{interestingness}, and \textit{confidence}.
% The validity of the score estimation approach lies in both the theoretical and empirical aspects.
\re{We consider the scores estimated by LLMs to be a valid proxy for humans' perception of meta relations from two perspectives.
First, most of the LLMs are trained to align with humans using various approaches, such as reinforcement learning from human feedback (RLHF)~\cite{wang2023aligning}.
Furthermore, previous experiments have verified that LLMs often make similar judgments to humans~\cite{zheng2024judging} and can simulate humans in HCI experiments~\cite{hamalainen2023evaluating}.}
\textit{Strength} provides a direct estimation of how strong the relation is.
Inspired by the metrics of evaluating fact strength~\cite{wang2019datashot}, \textit{fidelity} evaluates how much the relation covers the domain knowledge, and \textit{interestingness} implies whether the relation might attract story audiences.
Furthermore, we also ask LLMs to consider how much the relation can help convey the narrative intent via providing a score of \textit{helpfulness}.
Finally, to reflect how reliable the result is, a \textit{confidence} score is also returned by LLMs.
All scores are returned as an integer between 1 (the lowest) to 5 (the highest), similar to the 5-point Likert scale.
The final relation score is then computed based on these five scores as
$score = confidence * \sum (w_i * score_i)$, where $i \in \{strength, fidelity, helpfulness, interestingness\}$.
\re{The purpose of using a weighted average of these scores is to allow users to adjust the score definition based on their requirements.
In our experiments, we set all $w_i$s to 1 to treat all scores equally important.}
% when resolving the LLM outputs.
% In this way, we can suggest 
% The relation score can help users to 

Beyond the first task to return meta relations directly, the other two tasks are designed to facilitate the meta relation verification.
The second task is to \textit{facilitate the automatic verification of the identified meta relations}.
The LLMs are instructed to provide meta relations' detailed entities, which can be part of the data facts, such as attributes, filters, or focuses.
The task design is based on our observation that meta relations may not cover all perspectives of the fact.
For example, the meta relation in Fig.~\ref{fig:motivating_example} is mainly about the focuses of the data subspaces, \ie, ``Model = Toyota Prius'' and ``Model = Nissan Leaf, etc'' belong to different types of electric cars.
We ask LLMs to explicitly provide the entity information so that we can programmatically examine whether the meta relation is a valid relation about two facts by checking the existence of entities in facts.
When testing the module, we noticed that such information successfully filtered out problematic meta relations, such as relations between a domain knowledge point and a data fact.
% More details about resolving entities and filtering meta relations will be introduced later.

The last task is to \textit{facilitate the user's verification with eyeballing}.
Though we have designed the second task for automatic verification, it can only handle simple issues, such as the confusion between domain knowledge and data facts.
More complex issues like semantic errors in meta relations still rely on tool users to verify and resolve.
Therefore, we instruct LLMs to provide evidence about their suggested meta relations from two perspectives.
First, the LLM provides the original text in user-provide domain knowledge where the meta relation is derived.
In this way, whenever the user feels unsure about the meta relation, they can simply refer to the original text for verification.
Second, we also ask LLMs to explain how the relation is related to the narrative intent.
It can serve as a hint for users to decide whether the relation should be considered in the story.

The three tasks are finally described with natural language in our prompt, together with other necessary information, including input data, meta relation definition, and rules for output (\eg, output format).
Due to the limited space, we include the complete prompt in our supplementary material.
The whole prompt is then fed into LLMs for their responses.
After receiving the responses from LLMs, we conduct a series of output processing, including output format checking, automatic meta relation verification by examining the existence of entities in facts, and meta relation ranking with importance scores.

\subsection{Story Organization}\label{sec:story_organization}
Informed by the advantages of closely connected data analysis and storytelling~\cite{li2023notable}, we would like to design \tool as a notebook extension to organize data stories along with the progress of data analysis.
Therefore, the overall goal of this module is to find a suitable position for the newly added fact following its meta and data relations with previous facts.
The narrative intent is also considered in the organization (\textbf{C4}).
% Whenever the user selects a data fact through the analysis panel, it is automatically sent to the story piece organization module to find a suitable position for the corresponding data fact 
In this module, we apply LLMs in the story organization to cater to the needs of understanding the semantic meaning of meta relations and narrative intent (\textbf{C7}).
Similar to the meta relation identification module, we have a main task for inserting facts into slide decks and an auxiliary task for quick verification of results.
Both tasks leverage the same set of inputs, including the organized slide deck, the new fact, its data and meta relations with existing facts, and the narrative intent.

\begin{figure}
    \centering
    \includegraphics{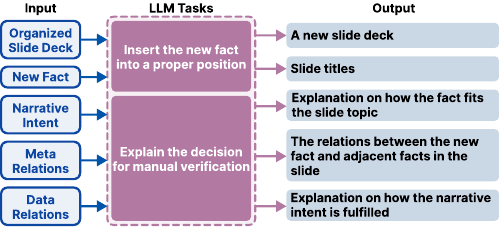}
    \caption{The figure explains the data story organization module.}
    \label{fig:organization_llm}
\end{figure}

The first task of \textit{inserting the fact into the slide deck} requires LLMs to consider the new fact's suitable position in the slide deck from the semantic and design levels.
From the semantic perspective, the model is instructed to consider how the fact fits the topic of the slide and the relations between the new fact and the facts behind or in front of it.
The narrative intent of the entire story should also be carefully considered.
At the slide design level, we provide two considerations, including the maximum number of facts in each slide to avoid overwhelming the story audience and the minimum alternation of the previously arranged sequence to diminish tool users' confusion.
Furthermore, a slide title is generated for each slide to summarize the content.
% Next, Lines 15-24 introduce the detailed output required from LLMs.
The second task requires LLM to \textit{provide the rationale behind the decision to insert the fact}.
The rationale should also be provided to explain the four design considerations at the semantic level.
In this task, we do not ask the LLM model to provide design alternatives, as multiple alternatives will increase users' workload in reading multiple stories and selecting their preferred one.
Considering that the goal of the organization module is to create a story draft during data analysis, we prefer not to increase users' mental load in organizing story pieces.
Another advantage is to reduce the LLM delay and cost.

% Finally, Line 25 emphasizes the rules that LLMs should follow, including the constraint of information usage and the output format.
The three tasks are then encapsulated into a prompt with the input data and requirements of output.
The complete prompt is available in our supplemental material.
After receiving LLM responses to the prompt, the module further processes the output.
Besides sanity checking, \tool ensures LLMs do not modify user-revised content, including the fact sequences and the slide title, in this step (\textbf{C5}).

\begin{figure*}
    \centering
    \includegraphics[width=\linewidth]{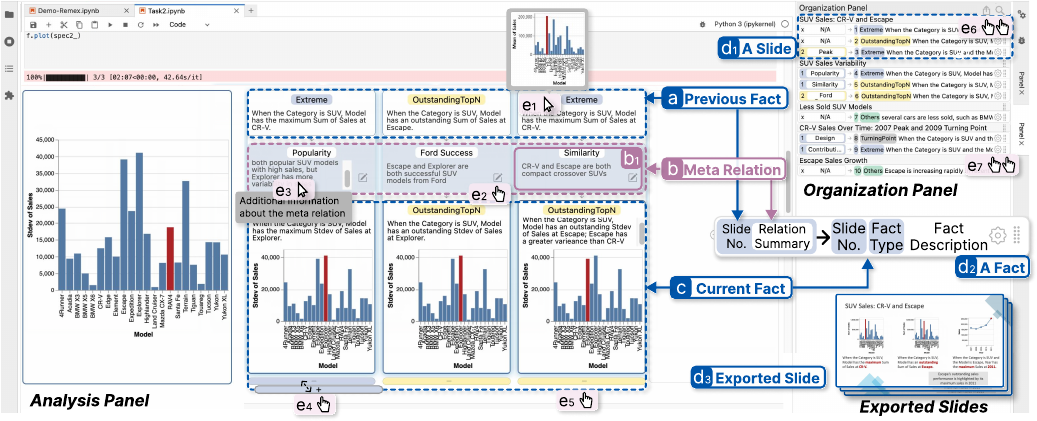}
    \caption{This figure shows \tool in JupyterLab with a case in the user study. \tool consists of multiple analysis panels under code cells and an organization panel to show how the data findings are organized into a sequence of slides. \re{(a)-(c) show previous facts, meta relations, and current facts in both analysis and organization panels. The arrows with (a)-(c) indicate that two linked areas show the same information. (d1) and (d2) explain a slide and a fact in the slide in the organization panel. (d3) shows exported slides. (e1)-(e7) explain the interactions.
    The arrow cursor indicates a hover interaction, while the hand cursor indicates a click interaction.}}
    \label{fig:interface}
\end{figure*}
% \vspace{-1em}
% \input{sections/04-01-computational-v1}
\subsection{Interactive Modules}\label{sec:interactive_modules}
To facilitate users' inspection and modification of outputs from computational modules (\textbf{C5}), 
\tool provides two interactive modules, including analysis panels presenting data facts and their meta relations to previously selected facts and an organization panel to show the sequence of all selected facts.
% This section introduces interactive modules.

% \begin{figure}
%     \centering
%     \includegraphics[width=0.7\linewidth]{figures/analysis.pdf}
%     \caption{This figure shows two relations in the analysis panel.}
%     \label{fig:interface}
% \end{figure}

\textbf{Analysis panel.} The analysis panel, depicted on the left side of Fig.~\ref{fig:interface}, displays the original chart on the left and accompanying data facts on the right. Users can select from recommended options or customize their facts.
Each data fact is presented in three cards.
The top card (Fig.~\ref{fig:interface}(a)) showcases the previous fact along with its type and description. Hovering over this card reveals the associated chart (Fig.~\ref{fig:interface}(e1)).
The middle card (Fig.~\ref{fig:interface}(b)) illustrates the meta relation between the previous and current facts, providing a summary word and detailed description. Users can edit this description via the edit button \re{(Fig.~\ref{fig:interface}(e2))}, with additional information for verification (see Sec.~\ref{sec:meta_identification}) available upon hovering \re{(Fig.~\ref{fig:interface}(e3))}.
The bottom card (Fig.~\ref{fig:interface}(c)) displays the current fact, including its type, description, and accompanying chart. For customized facts, users can choose from recommended options.
Later users can add/remove facts as needed (Fig.~\ref{fig:interface}(e4)-(e5)).

\textbf{Organization panel.}
The organization panel, depicted on the right side of Fig.~\ref{fig:interface}, enables users to outline and customize their stories. 
Selected data facts are displayed here, along with explanations for their placement in the dialog box (Fig.~\ref{fig:interface}(e4)).
In the panel, we represent each slide as a block (Fig.~\ref{fig:interface}(d1)) and individual facts as list items.
% Each block represents a slide (Fig.~\ref{fig:interface}(d1)), with list items encoding individual facts.
% \re{The information for previous facts is displayed on the left side of the block (Fig.~\ref{fig:interface}(a)-(b) with downward arrows), including its index and relation summary.}
% The layout (Fig.~\ref{fig:interface}(a)-(b)) shows the previous fact's information on the left
\re{The encoding of a list item is introduced with a legend (Fig.~\ref{fig:interface}(d2)). 
The left end of a list item indicates that a previously selected fact is linked with the new fact with a meta relation.
It shows a summary of the meta relation and the index of the slide where the previous fact is.}
The color encodes the previous fact's type for easy reference.
The current fact's details are on the right, including the index, fact type, and a text description of the fact. 
To export slides, users can simply click the icon as shown in Fig.~\ref{fig:interface}(e6).
The slide in Fig.~\ref{fig:interface}(d1) will be exported to a slide as Fig.~\ref{fig:interface}(d3) shows.
In Fig.~\ref{fig:interface}(e6), users can also hover on the magnifier icon to check the information for verifying story organization (see Sec.~\ref{sec:story_organization}).
Users can delete or reorder facts using buttons (Fig.~\ref{fig:interface}(e7)). 

\subsection{Slide Generation}
% When the sequence of facts is satisfactory, the user can export these facts as a slide deck.
\re{This module generates a slide deck based on the organized fact sequence.
Following Notable~\cite{li2023notable}, there are two styles of slides to accommodate facts from the same chart or different charts.
Furthermore, consider that meta relations can be a strong bond between two facts, we add an additional text box to include meta relations about two consecutive facts (see Fig.~\ref{fig:interface}(d2)).
We also provide the flexibility for users to customize slide background using a pre-defined set of backgrounds or through a link to the background image.}

\subsection{Implementation}
In \tool, the meta relation identification and the story organization modules are supported by an LLM.
After comparing the performance of \texttt{GPT-4-0613}~\cite{openai2023gpt4} and \texttt{GPT-3.5-Turbo}~\cite{gpt3.5} provided by Microsoft Azure, we adopted the standard version of \texttt{GPT-4-0613} with $temperature=0$ inside the two modules.
\re{Notably, our approach is not limited to a specific LLM.
We used the GPT family since they perform well~\cite{chiang2024chatbot} and are easy to access through Microsoft Azure APIs.}
% We selected \texttt{GPT-4}\cite{openai2023gpt4} provided by Microsoft Azure after comparing the performance of \texttt{GPT-4-0613} and \texttt{GPT-3.5-Turbo}~\cite{gpt3.5}.

% Though \texttt{GPT-3.5-Turbo} demonstrates its advantage in lower delay, it often fails to generate satisfactory responses, such as incorrect output format and confusion about background knowledge and previous data facts.
% Therefore, we finally adopted the standard version of \texttt{GPT-4-0613} with $temperature=0$ inside the two modules about meta relations.
% \input{sections/03-design-requirements}
% \input{sections/overview}
% \input{sections/04-approach}
\section{Exploratory Study}
% \tool serves as a proof-of-concept tool for 
To explore where and how AI should provide assistance to humans with meta relations, we conducted a user study to compare \tool and a recent notebook tool without considering meta relations~\cite{li2023notable}.

% To evaluate the effectiveness of \tool, we conducted a comparative user study with a recent data storytelling tool, Notable~\cite{li2023notable}.
% We consider Notable a comparable baseline tool since it also supports data storytelling in computational notebooks but only supports composing story pieces with considering data relations between them.
% Through comparing \tool with Notable, we hope to understand whether considering meta relations can enhance AI-powered tools in creating satisfactory data stories.

% This section presents an overview of \tool, a computational notebook tool that enables meta relation-based data storytelling.

% \subsection{Preliminaries}
% Before introducing the tool, we first illustrate how we define story pieces and relations between them in \tool.

\subsection{Preparation}
We designed a with-in subject user study to let participants have a hands-on experience with these two tools.

\textbf{Task and Dataset.} 
In our user study, each participant was instructed to finish two tasks with different tools in a counterbalanced manner.
To control the variables in the user study and minimize the memory effect, we design two tasks using two subsets from the same car sales dataset in previous research~\cite{wang2019datashot}.
Specifically, the participants were asked to create two stories to compare different car model sales in the \textit{Sedan} and \textit{SUV} categories with around eight data facts.
The dataset describes car sales between 2007 and 2011 with five attributes: \textit{model}, \textit{brand}, \textit{category}, \textit{sales}, and \textit{year}.
% To control the variables in the user study and minimize the memory effect, we split the dataset into two parts using the car categories so the participants could finish the same task with similar datasets and different tools.
% Furthermore, to ensure the fin
% Following previous research~\cite{li2023notable}, we cleaned the dataset in advance (\eg, removing all empty values) and reduced the size of the original dataset to five attributes covering common data types, including temporal, nominal, and quantitative data.
% Finally, the first subset contains four car categories (SUV, MPV, Pickup, and Sporty), while the other subset has another four (Subcompact, Compact, Midsize, and Fullsize).
% XXX movies between 19XX and 19XX, and the second includes XXX movies between 20XX and 20XX.
Considering the time cost of collecting domain knowledge, we provided participants with introductions to these car models from Wikipedia.
The introductions often briefly describe the cars' features and histories.
We also explained glossaries in car models, such as SUV and MPV, to help users understand the domain knowledge.
Besides, we prepared the \textit{cars} dataset~\cite{Vega} for participants to get familiar with two tools.

\textbf{Participants.}
\re{We invited 10 participants (5 males, 3 females, 2 prefer not to disclose, $Age_{mean}=25.5$, $Age_{std}=1.58$, denoted as U1-U10) from various disciplines, including computer science (U1, U4, U5, U8, U9), data science (U2, U6), social computing (U7), and business (U3, U10), with data analysis and communication backgrounds ($Experience_{mean}=5.30~years$, $Experience_{std}=2.06~years$) through our professional network.
Similar to our formative study, we invited participants with various expertise for a more comprehensive view of our research question.}
Each participant received \$12.5 as compensation after the user study.

\subsection{Procedure}
The whole in-lab user study lasts about 100 minutes in total.
After introducing the study procedure, we first collected participants' consent to record their videos and use their comments and demographic information for research purposes.
Next, we conducted a tutorial session for the first tool to help participants learn how to use it, taking around 10 minutes.
After the user felt confident about the tool usage, they started to use the first tool to compile a data story with assistance from the provided domain knowledge in around 30 minutes.
Then, we gave the tutorial of the second tool and asked them to finish the same task with another subset of data using 40 minutes in total.
% After the two sessions, the participants were asked to complete a questionnaire about the effectiveness and usability of these two tools.
% The usability questions follow the commonly applied System Usability Scale (SUS).
% The other questions about the tool effectiveness is available in Fig.\ref{}.
% where Q1-QX are about the effectiveness, and QX-QX evaluates the usability of these two tools following 
Finally, we interviewed the study participants about (1) how they perceive the differences between two tools brought by the introduction of meta relations; (2) how they leverage meta relations in the user study; (3) whether the approach of integrating AI for meta relations meet their common practice and expectations; and (4) how the tool can be further improved.
\re{After finishing all interviews, the results were summarized following a similar procedure as the formative study.}

% Participants: 12-16 data workers from the academia and the industry with experience in using notebooks

% Task: Finish data analysis and make data stories to present the results

% Stimuli:
% Two data analysis task (A and B) with a background information introduction (maybe from Wikipedia) and analytical question.
% Two corresponding notebook skeleton (only with data and package import code)
% Example: (Titanic analysis)

% \textbf{Procedure}
% Baseline: Notable + Wikipedia + ChatGPT
% Procedure:
% Collect background information, such as their job nature and their experience in authoring data stories (5-10 min)
% Session 1: Use Notable to finish the task A (10-min demo and 30-min task)
% Session 2: Use Notable+ to finish the task B (10-min demo and 30-min task)
% (The sequence of sessions 1 and 2 and the usage of datasets will be counterbalanced)
% Finish a questionnaire to compare two tools and an interview about the user experience (10-15 min)

% We will only ask them to select fact and organize instead of starting from Vega-Lite.

\subsection{Results}\label{sec:user_study_results}
% This section introduces the results of our user study.
% \subsubsection{Quantitative Results}

% \subsubsection{Qualitiative Results}
From the interview, we collected various opinions about meta relations.

\textbf{Integrating meta relations into the storytelling workflow demonstrates advantages from multiple perspectives.}
When asking the participants about the comparison between \tool and Notable, they generally considered that meta relations could help them with data analysis and the follow-up communication in notebooks.

We noticed that a widely recognized advantage brought by suggested meta relations is to provide context information for an in-depth understanding of data facts (U4, U5, U6, U7, U9).
For example, U5 commented that \tool presents implicit information in a visible way and therefore is more \textit{``insightful''}.
Meta relation can also help participants \textit{decide the directions of data analysis and communication} (U3, U5, U7, U8, U9).
U7 mentioned that such meta relations could suggest data facts that are potentially related to the focus of the current data facts from \textit{``high-level''} perspectives.
U7 pointed out a case to support his opinion.
As shown in Fig.~\ref{fig:interface}(b1), where \tool presents that ``CR-V and Escape are both compact crossover SUVs''.
Even though U7 was not familiar with these car models, it is possible to quickly decide that the sales trends of these two models may be worth more exploration.
U7 believed that there might be \re{some interesting sales trends behind these two compact crossover SUV models} since Escape has a large variance while CR-V has a larger overall sales.
Similarly, the meta relation ``Escape and Explorer are both successful SUV models from Ford'' on the left also stimulated U7 to explore more about their sales trends for more details.
Furthermore, U3 and U8 believed that \tool serves as an \textit{``external assistant''} (U3) to help them expand the angles of data exploration with meta relations.
U3 commented that \textit{``it provides a different viewpoint to help me break the limitation of my mindset''}.
Though agreeing with the benefit, U9 held a reservation about whether the benefit could surpass the potential risk of \textit{``converging too fast in data exploration''}.
U9 worried that some important data facts might not be explored due to the bias led by the suggested meta relations.

% assist in identifying the most important ones since it is hard to notice all relations among many facts.

Another advantage of meta relations, as expected, is that it provides a better connection between story pieces (U5, U6, U8, U9, U10).
U5 believed that the ways of organizing story pieces are enriched by introducing meta relations.
U9 commented that considering meta relations could make data stories more organized by connecting data facts more closely.
If only considering data relations, the story can be \textit{``fragmented''}.
The opinion was echoed by U8 and U10.

Besides, U8 also noticed that these relations, together with previous facts, could remind them of previously explored but forgotten findings.
U10 considered that the explicit suggestion of meta relations could 
reduce the efforts of considering them mentally, especially when \textit{``the relations between many facts form a network''}.

\textbf{The expected usage scenarios of meta relations are diverse.}
In \tool, we provide an approach to suggest meta relations with LLMs between user-interested data facts and newly identified data facts along with their data analysis.
Users can select suggested meta relations or manually add ones to share their opinions about how data facts are connected with AI for follow-up story organization.
% We hope to give users an opportunity to try one potential approach of considering meta relations with AI in data storytelling and ignite their ideas about expected approaches.
Though our current approach is appreciated for advantages like reminding users of previous facts and suggesting potential directions of analysis and storytelling, other approaches were proposed by our participants as well.

U1 mentioned that they prefer to focus on data facts instead of relations when initially exploring data.
Therefore, it is not desired to directly see the recommended meta relations, which may require considerable effort to understand and increase their workload.
A more desired way is to \textit{provide a widget where two facts can be selected by users and then meta relations can be inferred and presented to users}.
U1 believed this approach could help users understand the in-depth connection between two facts beyond simple data relations while minimizing users' efforts by only showing what they expect.
The idea was echoed by U3 since U3's common workflow is to \textit{``browse all facts first like reading menus''} and then build connections between them to tell stories.
U7 also shared a similar opinion, stating that a more fine-grained interaction was expected.
U7 imagined that users could select a specific data point for querying meta relations with other facts.

U5 believed that meta relations should not be limited to connecting observed facts, including both previously selected facts and the facts from the current chart.
They could play a more important role in data storytelling workflow by \textit{suggesting other potentially interesting facts associated with observed ones}.
U5 took car sizes (\eg, compact or mid-size) as an example.
They hoped that it was possible to recommend facts in other sizes when a meta relation related to a specific car size.
In this way, users can gain a more comprehensive view of related facts.
Similarly, U2 would like \tool to use meta relations to recommend what users might analyze in the next step.

U7 further proposed that \textit{some fixed meta relations based on factual knowledge could be presented before data analysis} in notebooks (\eg, the meta relations between items in the datasets).
The same idea was expressed by U2 and U5 as well.
In this way, users may get opportunities to become familiar with the meta relations prior to data analysis rather than learning new meta relations continuously during data analysis.
Furthermore, as U2 pointed out, such a pre-processing step might help reduce the time cost of LLMs in the follow-up meta relation recommendation.

\textbf{The display of meta relations is expected to be simple and effective.}
% From the user study, another important finding is that t
% From the user study, we learned an important lesson about the display of meta relations.
Since meta relations connect two data facts, when displaying meta relations to users, they require users not only to understand the relations themselves but also to recall an existing fact and to understand a new data fact.
Such an issue adds considerable cognitive load to users during the analysis.
As a result, the display of meta relations needs to be carefully designed to reach a balance between information size and the extra workload for users to understand it.
In our study, several users feel satisfied with the current design of meta relation display in \tool (U6, U8, U9, U10).
For example, U10 commented \textit{``the combination of charts and texts is nice''}.
Besides, we collected multiple suggestions on the display of meta relations to facilitate understanding.

U1 preferred to show meta relations only when users query.
U4 considered the current design to be \textit{``unnecessarily cumbersome''} since users need to read every meta relation.
To address the issue, U4 believed that a potential approach was to provide better summarizations of data facts and meta relations to users.
In U4's opinion, the current one-word summary was not informative, while the complete meta relation was lengthy.
An ideal summarization of U4 would be using one short sentence to describe each fact and another one to show the meta relation between them.
Another alternative approach would be showing a synthesized chart that introduces the two facts and the meta relation between them at once.
U7 described another relation-centric approach.
When users plot new charts, only meta relations might be shown initially so that users can have a quick glance.
When meta relations interest users, they may expand the facts for a detailed inspection and decide if they will be included in the story.

\textbf{Compared to domain knowledge, narrative intent is more dynamic.}
We observed that users heavily relied on meta relations from domain knowledge, while the role of narrative intent was less notable.
At the beginning of the analysis, most of the users did not have a clear intent about the story they wanted to tell.
Their narrative intent is gradually built and changed after they learn more about data.
For example, U6 noticed that the top-selling sedan for multiple years remained consistent, but its competitors changed a lot.
Then he decided to introduce the sales of cars year by year.
U7 noticed that some car models obviously have a greater standard deviation in sales and then decided to focus on the trend of car sales to explain the significant sales variance over the years.
Though \tool allows users to update their narrative intent whenever they want, we noticed that our study participants did not use the function.
We consider the main reason might be that users find it cumbersome to update it every time the intent is changed.
For example, U2 stated that they preferred an AI to follow their intent implicitly during the usage.

% Expected output:
% Quantitative results to indicate the pros and cons of Notable+ 
% Qualitative feedback on the reasons of advantages and disadvantages and future improvements

\section{Discussion}
This section discusses the lessons we learned, the limitations of our work, and future work.

% provides discussions about lessons learned from our research and also limitations and future work.

\subsection{Lessons}
\re{We learned three important lessons throughout our research project about applying meta relations in AI-powered data storytelling.}

% , including the value of explicitly introducing meta relations to story creation, the diverse scenarios where AI should be introduced to facilitate meta relations, and the application of LLMs.

\textbf{Making invisible meta relation sensible by creators and AI.}
Our research was motivated by the wide application of meta relations in practices, such as the example from Bloomberg (Fig.~\ref{fig:motivating_example}). 
However, meta relations were not reflected in the design of AI-powered storytelling tools, which led to the gap between humans' expected data stories and those created by AI.
One potential reason is that diverse meta relations are often implicitly considered by data story creators, like other meta information in visualization~\cite{burns2022invisible}.
As a result, the importance of meta relations for composing data stories is often overlooked.
\re{In this paper, we conducted research to explore meta relations in data stories from several angles, including their categorization and characteristics in practice and the timing and approach of AI assistance.}
% We also proposed considerations for introducing meta relations into the storytelling workflow with AI assistance and explored the scenarios where AI should intervene.
% The feedback from our exploratory user study verifies the necessity of considering meta relations explicitly to the data storytelling workflow.
% These meta relations not only assist users in planning their data story directions but also provide evidence for AI to organize data stories automatically.
% As a result, the consideration of meta relations can benefit human-AI collaboration in data storytelling with a better alignment between them.

During our research, we also realized an unresolved challenge in making humans and AI aware of meta relations.
Between the two sources of meta relations, \ie, domain knowledge and narrative intent, story creators and AI are aware of meta relations from domain knowledge easily.
We consider the reasons to be two-fold.
First, \textit{domain knowledge is almost static and objective while narrative intent is dynamic and personalized}.
Therefore, the synchronization between humans and AI in domain knowledge is convenient without much cost.
It is sufficient to provide domain knowledge once before the collaborative storytelling between humans and AI.
However, communicating narrative intent would be more challenging.
Story creators need to continuously update their intent with AI.
% The intent can be vague, so describing it with natural language 
Furthermore, describing the vague intent with natural language when communicating with AI can be hard.
% What makes it more challenging is that humans have to find a way to describe their vague intent to AI.
We found that \tool's function of updating narrative intent has almost never been used in our user study.
Second, \textit{the meta relations led by domain knowledge are mostly directly reflected by the connection between two facts, but narrative intent may not have a strong impact on a specific meta relation.}
Instead, narrative intent may affect the importance of meta relations.
For example, when the narrative intent is ``telling a story about luxury cars'', a meta relation ``Mercedes-Benz S-Series and BMW 7-Series are often selected by millionaires'' should be more important than ``Mercedes-Benz S-Series and BMW 7-Series are both cars from German carmakers.''
If the intent is ``telling a story about European cars'', then the relation about German carmakers is more likely to fit the story.
As the case shows, narrative intent may not be directly reflected in the relation descriptions, making it less sensible.
We consider that future research should delve into the challenge led by narrative intent, including how to facilitate effective communication of narrative intent between human creators and AI and how to make the impact of narrative intent more visible through visual design.

\textbf{Designing tools for applying meta relations in personalized ways.}
Our formative and user study informed the importance of offering sufficient personalization when designing AI-powered storytelling tools for meta relations.
We realized three levels of personalization in tool designs: \textit{meta relation source}, \textit{usage scenario}, and \textit{display}.

% Our formative study reveals two common meta relation sources, whic
According to our observations, \textit{source} preference varies among story creators.
Some preferred to have strong and clear narrative intent before creating stories~(\eg, P3), while others constructed their narrative intent gradually by identifying relations between story pieces~(\eg, U6, U7).
Considering categories of domain knowledge, we noticed that
\re{some users might use factual knowledge more frequently, while others considered conceptual and procedural knowledge more important}.
% It is essential to recommend meta relations aligned with their preference to enhance usefulness.

After suggesting meta-relations based on users' mindsets, the suggestions should be delivered using four approaches in different \textit{usage scenarios}.
% We noticed that these approaches can be roughly categorized into three types.
The first approach is to recommend meta relations when two facts are fixed.
It requires the lowest load of understanding meta relations but users also lose opportunities to identify potentially interesting meta relations beyond the selected facts.
The second one is to fix one fact and recommend multiple potential facts and meta relations between them.
The approach can help users decide the next step of analysis or storytelling with the assistance of meta relations.
The drawback is that it requires considerable computation to decide the following facts due to the large number of facts in datasets.
One potential solution is to leverage the Monte Carlo Tree Search~\cite{shi2020calliope, wu2023socrates, sun2022erato}.
The third method is to find meta relations between previously selected facts and a group of user-focused facts, as \tool does.
This approach balances the aforementioned approaches from the computational cost perspective since the space of facts is bounded but not limited to only two.
Users can also get unexpected meta relations to inspire their story, as our participants suggest (see Sec.~\ref{sec:user_study_results}).
However, it indeed requires users to understand more meta relations than the first approach, which is less favorable by some participants.
The last approach is to show meta relations before analysis to provide background information to users, similar to a mind map of factual knowledge to guide storytelling.
% It is like providing a mind map of factual knowledge to guide storytelling.
However, since new data facts are spotted in data analysis, the approach may not cover rich meta relations between newly discovered facts.
We would like to encourage future research to explore other potential usage scenarios~(\eg, identifying facts with specific meta relations) and investigate how to combine multiple types of assistance based on personal preferences.

\re{Finally, even if we can suggest suitable meta relations in a desired scenario, the \textit{display} of meta relations is also worth careful consideration.}
According to our study, some users suffered from information overload while others did not.
Such a difference requires us to design interactive displays of meta relations to cater to different needs.

\textbf{Applying an effective combination of LLMs and rule-based approaches for considering meta relations.}
% The last lesson is to design a effective combination of advanced AI systems and rule-based approaches within 
Compared to data storytelling tools based on facts and relations from data~\cite{wang2019datashot, shi2020calliope}, the tools with meta relations require interpreting and suggesting natural language-based meta relations with LLMs.
% They require advanced LLMs for the meta relation-related tasks.
However, LLMs are not without limitations.
For example, their high delay has affected the user experience with \tool (U5, U8, U9, U10).
\re{To help readers understand the issue, we report the time spent identifying meta relations (Min: 25, Max: 80, Mean: 47, Std: 16) and organizing data facts (Min: 22, Max: 56, Mean: 37, Std: 13) in seconds with GPT-4 in U5's session.}
Sometimes the identified meta relations are unreasonable or not in-depth.
% \haotian{add some feedback from user study}
For example, U2 noticed that some meta relations returned by LLMs were only a repetition of two facts (\eg, a suggested meta relation between a fact about increasing sales of BMW and another indicating Toyota as the brand with highest sales is ``BMW Sales decrease while Toyota has the highest sales'').
U8 commented that the background information provided by meta relations was not out of expectation.
Furthermore, LLM's ability to understand and analyze data is limited~\cite{jin2023numgpt, cheng2023gpt, narayan2022can}.
% Therefore, they may not produce accurate data analysis results.
Their bias~\cite{ferrara2023should} and hallucinations~\cite{zhang2023siren} even damage the credibility. 

To address these issues, an effective combination of heuristics-based approaches with LLMs is adopted.
We consider heuristics-based approaches to be more deterministic but may not be as capable as LLMs. 
Therefore, they can be leveraged to enhance the reliability and accuracy of AI-powered tools, while LLMs can provide more flexibility.
For example, in \tool, we leverage rule-based approaches to compute data facts and apply LLMs to handle meta relations.
We also apply rule-based automatic verification for LLM-generated content, such as checking the integrity of the two facts connected by a meta relation.
Moreover, the combination of algorithms can be enhanced in future tools.
For example, the delay led by LLMs might be reduced by distilling meta relations from factual knowledge prior to the entire storytelling workflow or introducing external knowledge bases~\cite{wang2023open}.
% In this way, LLMs could focus on the meta relations about newly discovered data facts during analysis.

Beyond combining LLMs and rule-based approaches, a combination of LLMs and other deep learning methods is worth exploring.
For example, inspired by prior research~\cite{zheng2024judging, shankar2024spade}, we might employ another LLM to check the suggested meta relations and organized data story, considering the inability of rule-based methods.
% It can also further reduce users' manual effort for verification. 
Besides, U2 pointed out that a large quantity of domain knowledge could enhance the usefulness of \tool but worried about the scalability.
Retrieval-augmented generation (RAG)~\cite{lewis2020retrieval} might be applied to address the issue.

% We believe this lesson is essential for various data storytelling tools since data stories combine rigorous data facts and narrations to convey coherent and faithful messages from datasets~\cite{cheng2022investigating}.
% It is worth exploring how to combine different methods to achieve effective and efficient AI systems for human-AI collaborative data storytelling.

% \textbf{Applying meta relations as a medium to enhance the alignment between humans and AI}

% \textbf{Supporting flexible usage of meta relations in practices}

% \textbf{What are the pros and cons of considering flexible relations beyond simple data relations?}

% \textbf{How should we design an effective combination of LLM and rule-based approaches?}

% \textbf{How should we facilitate human-AI collaboration with humans' feedback to AI?}

\subsection{Limitations and Future Work}
% Our work describes a preliminary investigation into meta relations and how to facilitate the usage of meta relations with AI.
As a preliminary investigation,
we admit that our work has limitations and hope the understanding of meta relations can be enhanced in future.

The limitation of our research mainly lies in the design of our user study.
% The user study may not cover all situations in practice.
First, considering the time limit, we provided users with a relatively small dataset and asked them to create a short story following prior research~\cite{wang2019datashot, li2023notable}.
U7 believed that meta relations could have demonstrated their values more clearly when the dataset was more complex.
Future research should conduct more field studies or longitudinal user studies to observe the practices of meta relations for more insights.
\re{Second, the scale of our exploratory study was limited to ten data workers.
We believe it will be beneficial if the scale could be extended with the involvement of other potential users, such as data journalists and researchers in natural science.
A larger experiment can reveal more insights regarding where and how AI could support meta relation usages, such as more scenarios and interface design preferences.
We hope future research could investigate the topic with more participants to substantiate our findings.}
Third, \tool only covers limited cases of data storytelling.
\tool serves for telling stories with slideshows, one of the most common formats of data stories~\cite{hullman2013deeper}.
However, data storytelling covers a broad spectrum beyond slideshows~\cite{segel2010narrative},
such as telling stories as data articles~\cite{sultanum2021leveraging, latif2021kori} or even
in virtual reality or augmented reality environments~\cite{yang2023understanding, tong2024vistellar}.
% about scientific visualizations~\cite{liao2014storytelling, morth2022scrollyvis}.
It is interesting to explore whether our findings can be generalized.

We are also interested in various future directions to enhance our research.
First, we hope to further enhance the theoretical foundation of meta relations.
% For example, we propose to categorize meta relations based on their sources according to our formative study.
For example, it will be interesting to investigate approaches to categorize meta relations other than their sources and propose design spaces for future usage.
\re{Second, \tool can be enhanced from multiple perspectives.
For example, more formats of domain knowledge, \eg, figures, can be supported by \tool.}
It will also be interesting to enhance \tool's compatibility with other software, such as video editing tools, to support data storytellers' practices.

% develop software for applying meta relations in multiple environments.
% % Data storytellers may work with different software for data analysis to video editing.
% As our user study results imply, meta relations may have their roles in different stages, including analysis and planning.
% Therefore, it is essential to develop plugins to support the entire workflow of data storytellers.
% % with centralized management.

% \textbf{Tool design}

% \textbf{Evaluation}

% \textbf{}
\section{Conclusion}
Meta relations provide a unique perspective for data storytellers to connect various story pieces into a coherent and vivid story.
However, they have not been thoroughly examined and considered in AI-powered data storytelling, leaving a gap between AI-created data stories and humans' expectations.
To address the issue, we conducted a formative study to understand story creators' practices regarding meta relations, such as common meta relation sources and their pain points about meta relations.
Next, based on the design consideration learned from the formative study, we deepen the understanding of where and how AI should take actions in the storytelling workflow for meta relations through a user study with a proof-of-concept tool, \tool.
The user study revealed a series of findings about applying meta relations in the data storytelling workflow, including the diverse benefits and various expected usage scenarios.
As an initial step towards introducing meta relations into AI-powered storytelling tools, we also learned multiple precious lessons.
% , including the personalized usage of meta relations and the necessity of making meta relations explicit.
\re{We hope future research could enhance the theoretical foundation of meta relations and their applications in practice.}

\iffalse
\acknowledgments{
The authors wish to thank A, B, and C. This work was supported in part by
a grant from XYZ (\# 12345-67890).}
\fi
%\bibliographystyle{abbrv}
% \bibliographystyle{abbrv-doi}
%\bibliographystyle{abbrv-doi-narrow}
% \bibliographystyle{abbrv-doi-hyperref}
%\bibliographystyle{abbrv-doi-hyperref-narrow}
\bibliographystyle{IEEEtran}

\balance
\bibliography{references}

\begin{IEEEbiography}[{\includegraphics[width=1in,height=1.2in,clip,keepaspectratio]{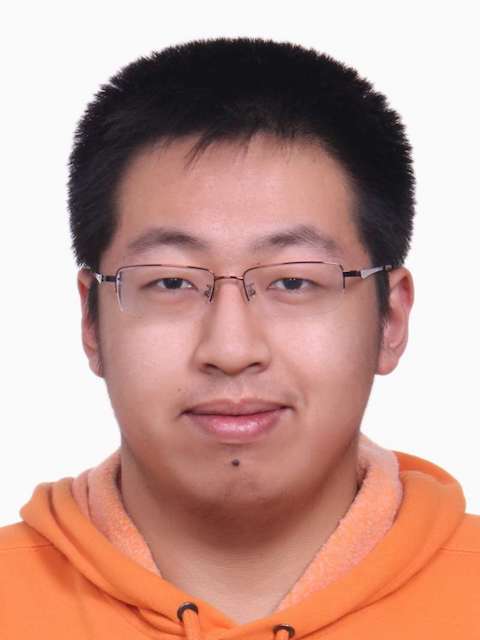}}]{Haotian Li}
is currently a Ph.D. candidate in Computer Science and Engineering at the Hong Kong University of Science and Technology (HKUST). His main research interests are data visualization, visual analytics, and human-computer interaction. He received his BEng in Computer Engineering from HKUST. For more details, please refer to \url{https://haotian-li.com/}. 
\end{IEEEbiography}

\begin{IEEEbiography}[{\includegraphics[width=1.0in,height=1.25in, clip,keepaspectratio]{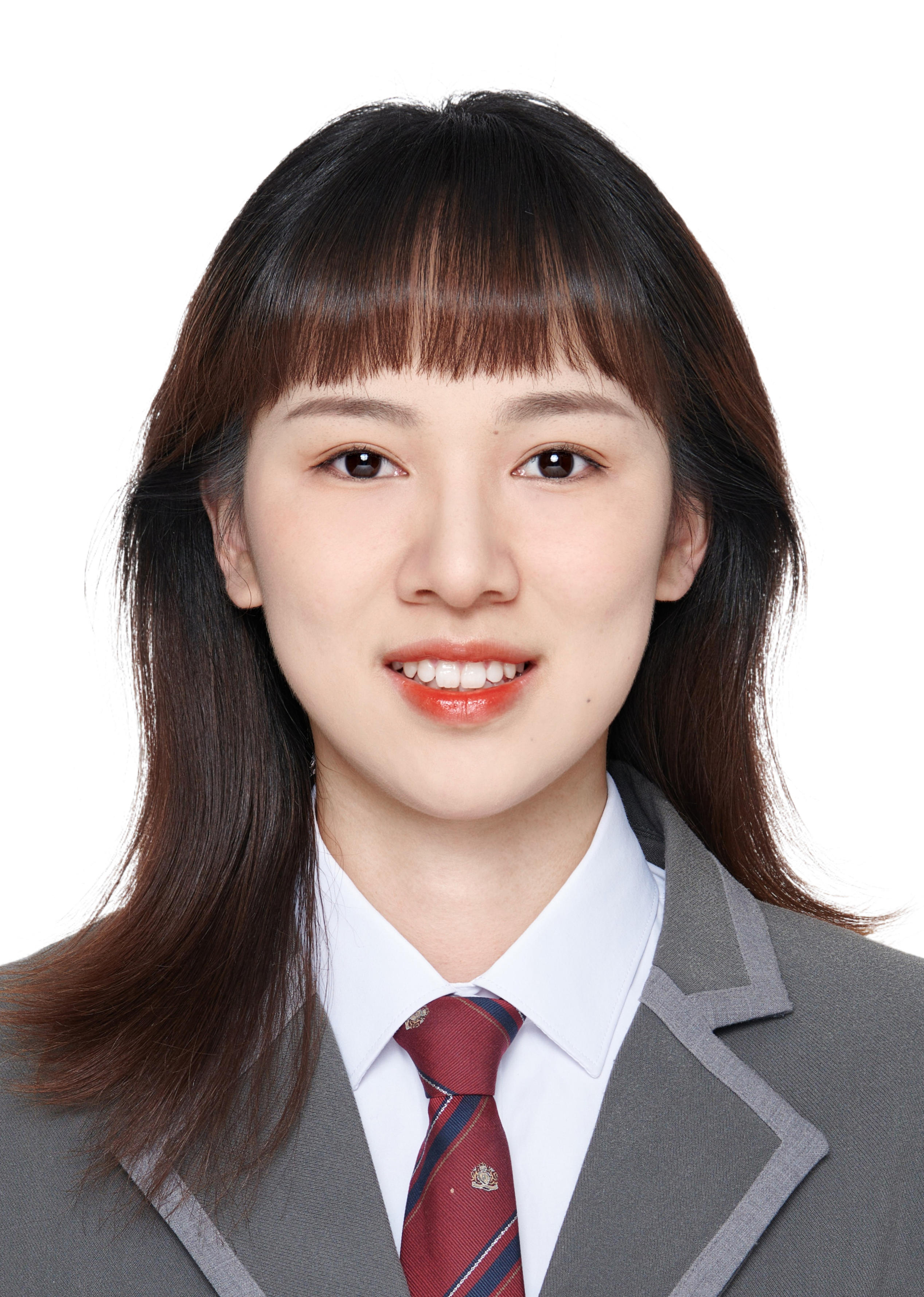}}]{Lu Ying} is currently a Ph.D. candidate at the State Key Lab of CAD\&CG, Zhejiang University. Her main research interests are on data storytelling, glyph-based visualization. She is dedicated to integrating the AI technique into visualization to ease the creation of visualization. She received her BEng in Digital Media Technology from Zhejiang University. For more details, please refer to \url{https://yiyinyinguu.github.io/}.
\end{IEEEbiography}

\begin{IEEEbiography}[{\includegraphics[width=1in,height=1.2in,clip,keepaspectratio]{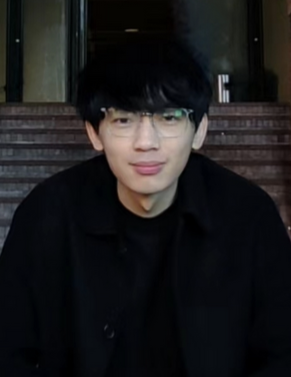}}]{Leixian Shen}
is a PhD student in the Department of Computer Science and Engineering at The Hong Kong University of Science and Technology. He received his master's degree in Software Engineering from Tsinghua University in 2023 and obtained his bachelor's degree in Software Engineering from Nanjing University of Posts and Telecommunications in 2020. His research interests include visual data analysis and storytelling. For more details, please refer to \url{https://shenleixian.github.io/}. 
\end{IEEEbiography}

\begin{IEEEbiography}[{\includegraphics[width=1in,height=1.2in,clip,keepaspectratio]{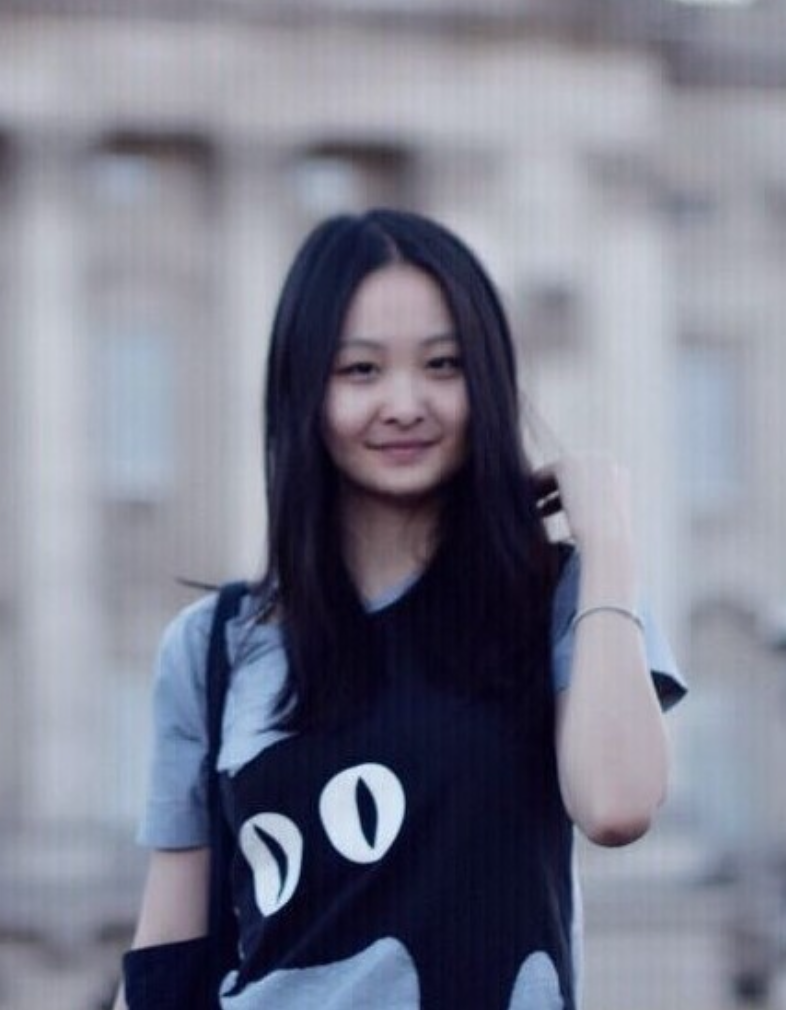}}]{Yun Wang}
is a senior researcher in the Data, Knowledge, Intelligence (DKI) Area at Microsoft. Her research lies in the intersection of Human-Computer Interaction, and Information Visualization. She develops innovative techniques and interactive systems to facilitate Human-AI Collaboration, Human-Data Interaction, Visual Communication, and Data Storytelling through an interdisciplinary approach. She received her Ph.D. from the Hong Kong University of Science and Technology. For more details, please refer to \url{https://www.microsoft.com/en-us/research/people/wangyun/}. 
\end{IEEEbiography}

\begin{IEEEbiography}[{\includegraphics[width=1in,height=1.25in,clip,keepaspectratio]{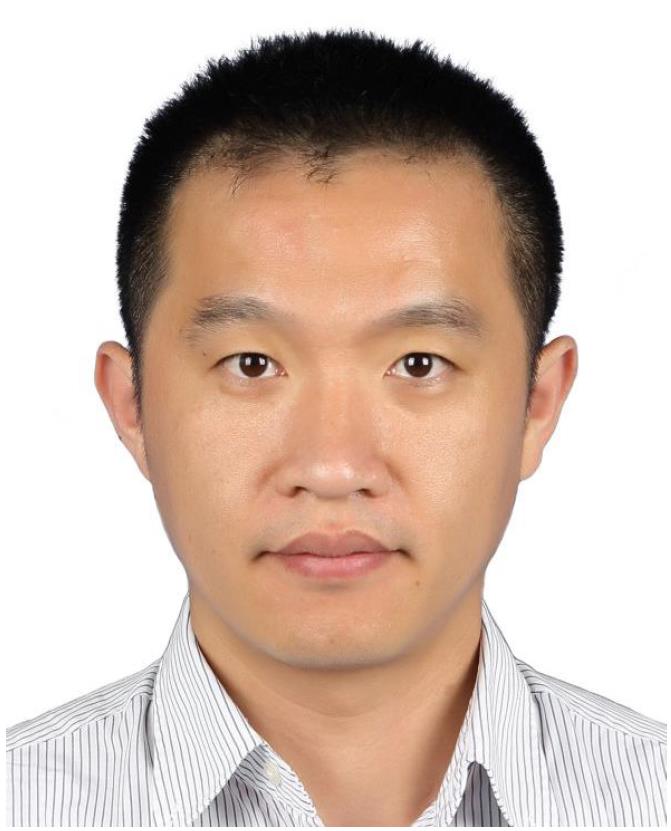}}]{Yingcai Wu} is a Professor at the State Key Lab of CAD\&CG, Zhejiang University.
His main research interests are information visualization and visual analytics, with focuses on urban computing, sports science, immersive visualization, and social media analysis. 
He received his Ph.D. degree in Computer Science from The Hong Kong University of Science and Technology. 
Prior to his current position, Dr. Wu was a postdoctoral researcher in the University of California, Davis from 2010 to 2012, and a researcher in Microsoft Research Asia from 2012 to 2015. 
For more information, please visit \url{http://www.ycwu.org}.
\end{IEEEbiography}

\begin{IEEEbiography}[{\includegraphics[width=1in,height=1.2in,clip,keepaspectratio]{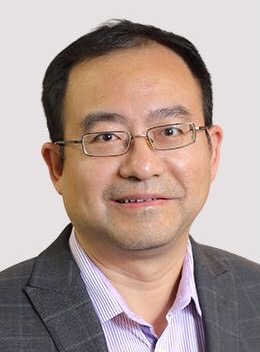}}]{Huamin Qu}
is a chair professor in the Department of Computer Science and Engineering (CSE) at the Hong Kong University of Science and Technology (HKUST) and also the director of the interdisciplinary program office (IPO) of HKUST. He obtained a BS in Mathematics from Xi'an Jiaotong University, China, an MS and a PhD in Computer Science from the Stony Brook University. His main research interests are in visualization and human-computer interaction, with focuses on urban informatics, social network analysis, E-learning, text visualization, and explainable artificial intelligence (XAI). For more information, please visit \url{http://huamin.org/}.
\end{IEEEbiography}

\end{document}